\documentclass[a4paper]{jpconf}
\usepackage{graphicx}

\begin{document}

\title{Multiple Lifshitz transitions driven by short-range antiferromagnetic
correlations in the two-dimensional Kondo lattice model}
\author{Yu Liu and Guang-Ming Zhang}

\address{State Key Laboratory of Low-Dimensional Quantum Physics and
Department of Physics, Tsinghua University, Beijing, 100084,
China}

\ead{gmzhang@tsinghua.edu.cn}

\begin{abstract}
With a mean field approach, the heavy Fermi liquid in the two-dimensional
Kondo lattice model is carefully considered in the presence of short-range
antiferromagnetic correlations. As the ratio of the local Heisenberg
superexchange coupling to the Kondo coupling increases, the Fermi surface
structure changes dramatically. From the analysis of the ground state energy
density, multiple Lifshitz type phase transitions occur at zero temperature.
\end{abstract}

\section{Introduction}

Quantum phase transitions have attracted much interest in studying
correlated electron systems. An electronic phase transition
associated with the change of Fermi surface (FS) topology, the
so-called Lifshitz transition\cite{Lifshitz-1960}, can be induced
without any spontaneous symmetry breaking and local order
parameter. The Kondo lattice model is a prototype model and
believed to capture the basic physics of heavy fermion materials.
The huge mass enhancement of the quasiparticles can be attributed
to the coherent superposition of individual Kondo screening
clouds, and the resulting metallic state is characterized by a
\textit{large} FS with the Luttinger volume containing both
conduction electrons and localized moments. Competing with the
Kondo singlet formation, the localized spins indirectly interact
with each other via magnetic polarization of the conduction
electrons -- the Ruderman-Kittel-Kasuya-Yosida interaction. Such
an interaction dominates at low values of the Kondo exchange
coupling and is the driving force for the
antiferromagnetic (AFM) long-range order and quantum phase transitions\cite%
{Doniach,zhang-2000}. So far most of investigations focus on the possible FS
reconstruction around the magnetic quantum critical point\cite%
{Si-2001,Senthil-2004,Ogata-2007,Assaad-2008,Senthil-2009}. However, the FS
topology in the paramagnetic heavy fermi liquid phase may also be
drastically changed by the short-range AFM spin correlations between the
localized spins, leading to the Lifshitz phase transitions\cite{zhang-su-yu}%
. The nature of such a quantum phase transition has not been thoroughly
explored yet.

In this paper, we consider the two-dimensional Kondo lattice model with the
Heisenberg AFM superexchange coupling between localized spins. By
introducing uniform short-range AFM valence-bond and Kondo screening
parameters, a fermionic mean-field theory is derived and carefully
re-examined. Away from half-filling, at the conduction electron density $%
n_{c}=0.85$, for example, the possible changes of FS topology in
the paramagnetic heavy Fermi liquid phase are considered carefully
as increasing the short-range AFM spin correlations.

\section{Mean field theory}

The model Hamiltonian defined on a square lattice is given by
\begin{equation}
H=\sum_{\mathbf{k},\sigma }\epsilon _{\mathbf{k}}c_{\mathbf{k}\sigma
}^{\dagger }c_{\mathbf{k}\sigma }+J_{K}\sum_{i}\mathbf{S}_{i}\cdot \mathbf{s}%
_{i}+J_{H}\sum_{\left\langle ij\right\rangle }\mathbf{S}_{i}\cdot \mathbf{S}%
_{j}.  \label{eqn1}
\end{equation}%
The spin-1/2 operators of the local magnetic moments have the fermionic
representation $\mathbf{S}_{i}=\frac{1}{2}\sum_{\sigma \sigma ^{\prime
}}f_{i\sigma }^{\dagger }\mathbf{\tau }_{\sigma \sigma ^{\prime }}f_{i\sigma
^{\prime }}$ with a local constraint $\sum_{\sigma }f_{i\sigma }^{\dagger
}f_{i\sigma }=1$, where $\mathbf{\tau }$ is the Pauli matrices. Following
the large-$N$ fermionic approach\cite{Coleman-1989}, the Kondo spin exchange
and Heisenberg superexchange terms can be expressed up to a chemical
potential shift as
\begin{equation}
\mathbf{S}_{i}\cdot \mathbf{S}_{j}=-\frac{1}{2}\sum_{\sigma \sigma ^{\prime
}}f_{i\sigma }^{\dagger }f_{j\sigma }f_{j\sigma ^{\prime }}^{\dagger
}f_{i\sigma ^{\prime }}, \hspace{1cm}\mathbf{S}_{i}\cdot \mathbf{s}_{j}=-%
\frac{1}{2}\sum_{\sigma \sigma ^{\prime }}f_{i\sigma }^{\dagger }c_{j\sigma
}c_{j\sigma ^{\prime }}^{\dagger }f_{i\sigma ^{\prime }},
\end{equation}%
then uniform short-range AFM valence bond and Kondo screening
order parameters can be introduced as
\begin{equation}
\chi =-\sum_{\sigma }\left\langle f_{i\sigma }^{\dagger }f_{i+l\sigma
}\right\rangle ,\hspace{1cm} V=\sum_{\sigma }\left\langle c_{i\sigma
}^{\dagger }f_{i\sigma }\right\rangle .  \label{eqn2}
\end{equation}

To avoid the incidental degeneracy of the conduction electron band on a
square lattice, we choose $\epsilon _{\mathbf{k}}=-2t\left( \cos k_{x}+\cos
k_{y}\right) +4t^{\prime }\cos k_{x}\cos k_{y}-\mu $, where $t$ and $%
t^{\prime }$ are the first and second nearest neighbor hoping matrix
elements, respectively, while $\mu $ is the chemical potential, which should
be determined self-consistently by the density of the conduction electrons $%
n_{c}$. Under the uniform mean-field approximation, the f-spinons form a
very narrow band with the dispersion $\chi _{\mathbf{k}}=J_{H}\chi \left(
\cos k_{x}+\cos k_{y}\right) +\lambda $, where $\lambda $ is the Lagrangian
multiplier to be used to impose the local constraint on average.

Thus the corresponding mean-field Hamiltonian reads
\begin{equation}
H=\sum_{\mathbf{k}\sigma }\left(
\begin{array}{cc}
c_{\mathbf{k}\sigma }^{\dagger } & f_{\mathbf{k}\sigma }^{\dagger }%
\end{array}%
\right) \left(
\begin{array}{cc}
\epsilon _{\mathbf{k}} & -\frac{1}{2}J_{K}V \\
-\frac{1}{2}J_{K}V & \chi _{\mathbf{k}}%
\end{array}%
\right) \left(
\begin{array}{c}
c_{\mathbf{k}\sigma } \\
f_{\mathbf{k}\sigma }%
\end{array}%
\right) +E_{0},
\end{equation}%
with $E_{0}=N\left( -\lambda +J_{H}\chi ^{2}+J_{K}V^{2}/2\right) $. The
quasiparticle excitation spectra can be easily obtained
\begin{equation}
E_{\mathbf{k}}^{\pm}=\frac{1}{2}\left[ \left( \epsilon _{\mathbf{k}}+\chi _{%
\mathbf{k}}\right) \pm W_{\mathbf{k}}\right] ,  \label{eqn3}
\end{equation}
which implies that the conduction electron band $\epsilon _{\mathbf{k}}$ has
a finite hybridization with the spinon band $\chi _{\mathbf{k}}$. Here $W_{%
\mathbf{k}}=\sqrt{\left( \varepsilon _{\mathbf{k}}-\chi _{\mathbf{k}}\right)
^{2}+\left( J_{K}V\right) ^{2}}$. Accordingly, the ground-state energy
density can be evaluated as
\begin{equation}
\varepsilon _{g}=\frac{2}{N}\sum_{\mathbf{k,\pm }}E_{\mathbf{k}}^{\pm}\theta
\left( -E_{\mathbf{k}}^{\pm}\right) -\lambda +J_{H}\chi ^{2}+\frac{1}{2}%
J_{K}V^{2},  \label{eqn4}
\end{equation}%
where $\theta ( -E_\mathbf{k} )$ is the theta function. Then the
self-consistent equations for the mean-field variables $\chi $, $V$, and $%
\lambda $ and the chemical potential $\mu $ can be deduced from
the relations
\begin{equation}
\frac{\partial \varepsilon _{g}}{\partial \chi }=0,\hspace{0.5cm} \frac{%
\partial \varepsilon _{g} }{\partial V}=0, \hspace{0.5cm} \frac{\partial
\varepsilon _{g}}{\partial \lambda }=0,\hspace{0.5cm}
n_{c}=-\frac{\partial \varepsilon _{g}}{\partial \mu. }
\end{equation}

\section{Results and Discussions}

In the following, we will assume that $t^{\prime }/t=0.1$ and $n_{c}=0.85$.
When the self-consistent equations are carefully evaluated, we find that the
mean-field AFM order parameter $\chi $ is \textit{always} positive in the
range $0<J_{H}/J_{K}\leq 3$ so that the resulting state is a stable
paramagnetic metal.

In Fig.1, as the strength of the AFM spin fluctuations grows up, we present
the evolution of the band structure of the renormalized heavy quasiparticles
around the Fermi level in the direction $(0,0)\rightarrow (\pi ,\pi
)\rightarrow (\pi ,0)\longrightarrow (0,0)$ of the first Brillouin zone.
Fig.1a to Fig.1h correspond to $J_{K}/t=2.0$ and $x=J_{H}/J_{K}=0$, $0.11$, $%
0.12$, $0.234$, $0.235$, $1.2$, $2.76$, and $2.77$, respectively.
At the critical values $x=0.1181$, the point $(\pi ,\pi )$ changes
from the local maximum to local minimum of the quasiparticle band,
while at $x=0.2341$, the separated band around the momentum $(\pi
,\pi /2)$ starts to move away from the Fermi level. As the ratio
of the coupling strengths increases, the separated
band moves to the momentum $(\pi ,0)$, and reaches the Fermi level again at $%
x=2.77$.

\begin{figure}[h]
\begin{minipage}{18pc}
\includegraphics[width=18pc]{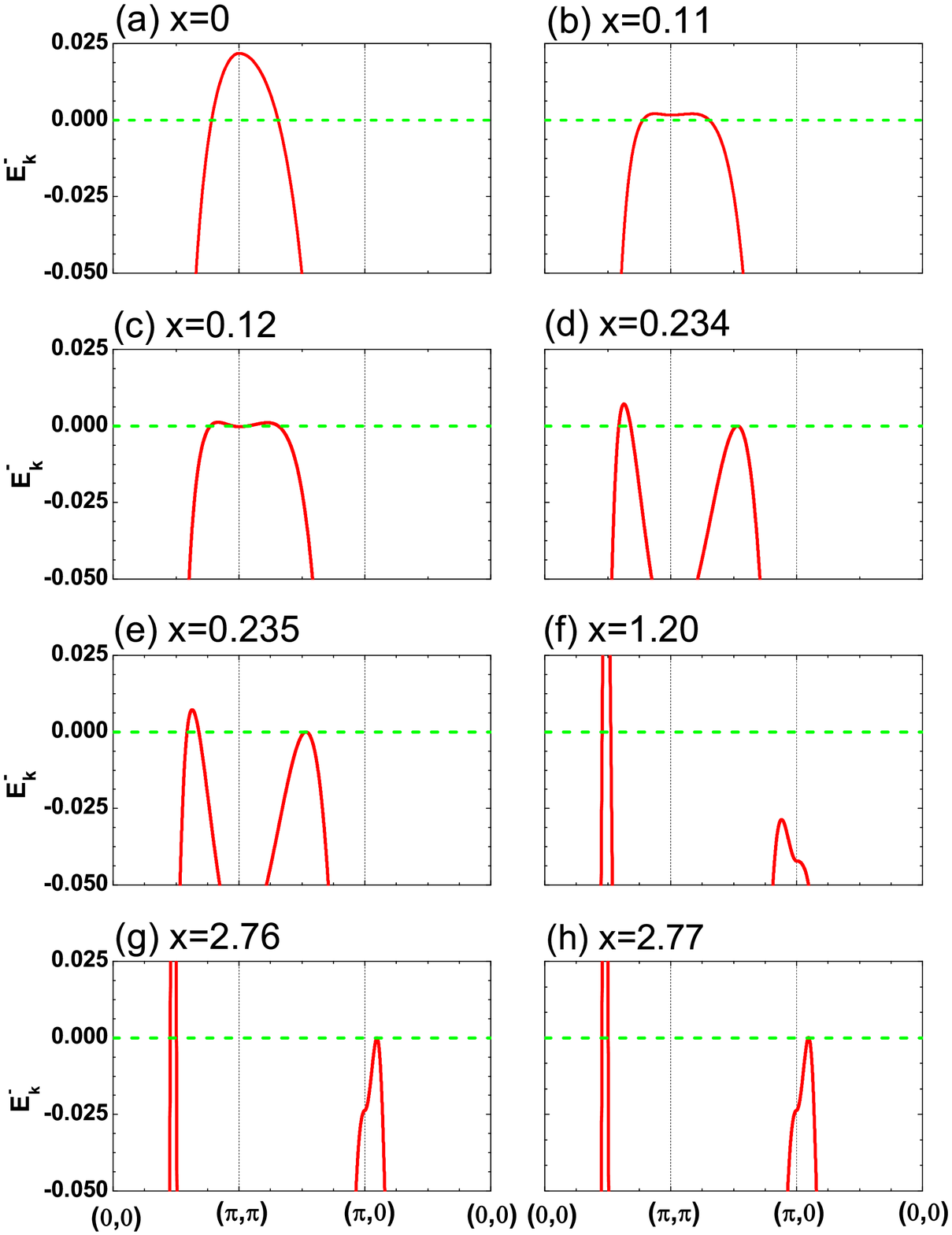}
\caption{The lower renormalized quasiparticle band in the
direction $(0,0)\rightarrow (\pi,\pi)\rightarrow
(\pi,0)\rightarrow (0,0)$ as increasing the strength of the AFM
spin fluctuations.}
\end{minipage}\hspace{2pc}
\begin{minipage}{18pc}
\includegraphics[width=18pc]{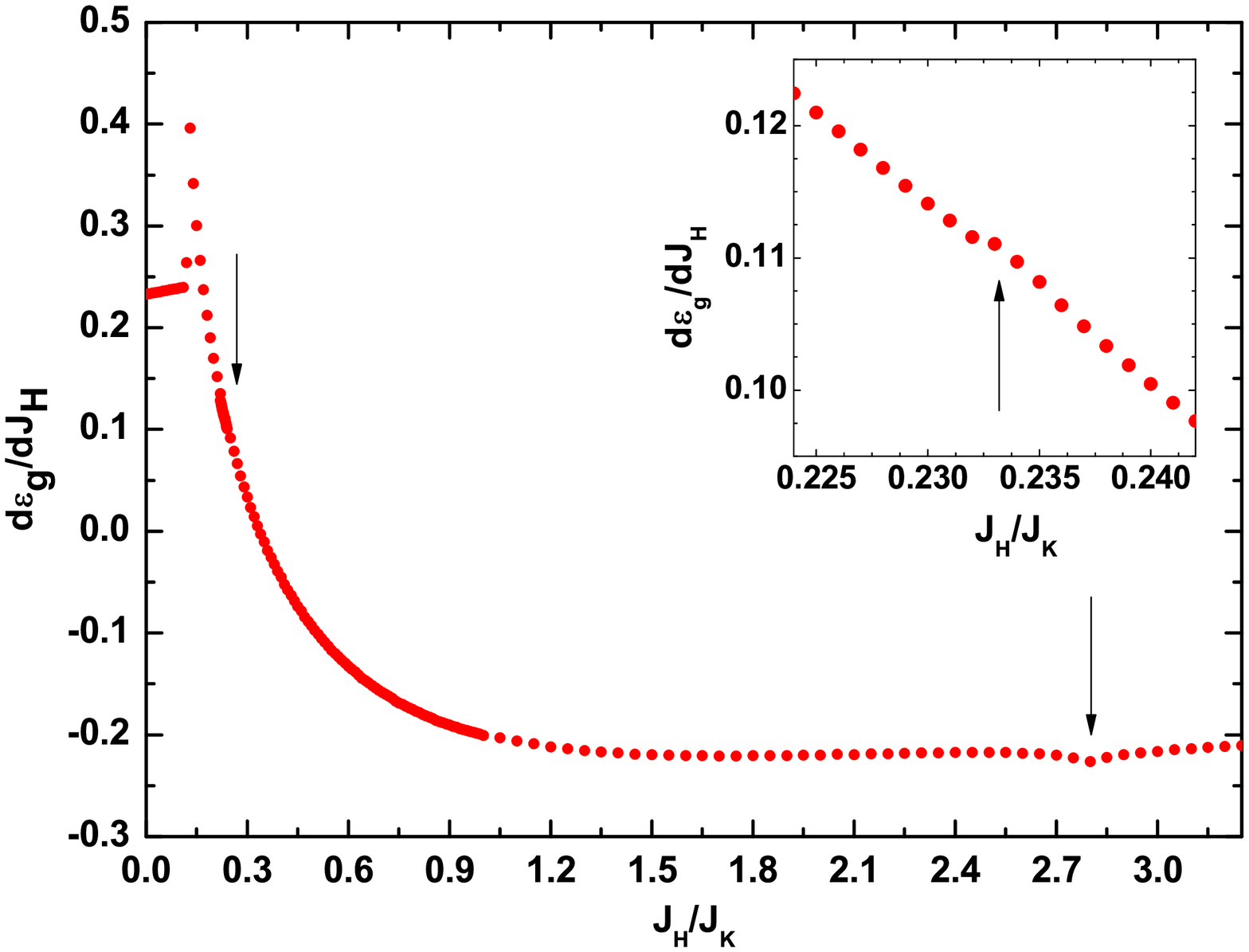}
\caption{The first-order derivative of the ground state energy as
a function of the parameter $x=J_{H}/J_{K}$. There are three
singular points, $x_{c,1}=0.1181$, $x_{c,2}=0.2341$, and
$x_{c,3}=2.77$. The first one corresponds to a first order
transition, while the latter two correspond to two second order
phase transitions, respectively.}
\end{minipage}
\end{figure}

From the quantum phase transition aspects, we calculate the ground state
energy density $\varepsilon _{g}$ and its first-order derivative with
respect to the ratio of the coupling parameters $x$. The numerical results
are displayed in Fig.2. We find that there are three non-analytical points. $%
\varepsilon _{g}$ is finite and continuous in the parameter range $0<x<3$,
However, its first-order derivative has a large jump at $x_{1c}=0.1181$,
corresponding to a first-order quantum phase transition. Moreover, two small
kinks appear at $x_{2c}=0.2341$ and $x_{3c}=2.77$ in the first-order
derivative, which correspond to the jumps in the second-order derivative of $%
\varepsilon _{g}$. So $x_{2c}$ and $x_{3c} $ denote two second-order quantum
phase transitions.

Once the renormalized quasiparticle band structure is available,
the corresponding FS can be easily obtained. The obtained FS is
the large hole-like one and shown as the shaded area. In
Fig.3a-3g, the center of the FS is shifted from ($0,0$) to ($\pi
,\pi $). For a fixed $J_{K}/t=2.0$, we can see that the FS is a
hole-like circle around ($\pi ,\pi $) for the parameter range
$0\leq x \leq 0.11$. At $x_{1c}=0.1181$, the topology of the FS
starts to change: a small circle emerges in the center of the
deformed large square FS. As $x$ is further increased, both
circles expand and the small one is deformed into a rotated
square. Up to $x_{2c}=0.2341$, the two deformed circles intersect
each other and then decompose into four Fermi pockets. When x
increases to $x_{3c}=2.77$, the FS pockets are reconnected again
and form two closed square-like circles. Three critical values
correspond to three different quantum phase transitions, which
belong to the category of Lifshitz phase transitions.

\begin{figure}[tbp]
\center{\includegraphics[width=30pc]{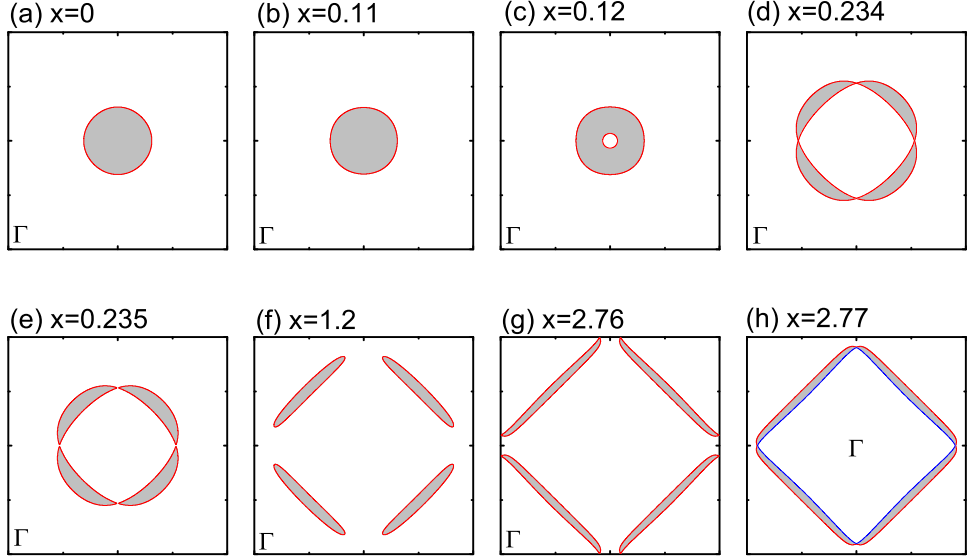}}
\caption{The structure of the FS changes as increasing the
strength of the AFM spin fluctuations. The shaded area represents
the hole-like FS. The center of the FS is shifted at $(\pi,\pi)$
in (a)-(g).}
\end{figure}

To some extent our present mean-field theory captures the heavy-fermion
liquid physics of the Kondo-Heisenberg lattice systems, especially the Fermi
surface evolution of the renormalized heavy quasiparticles as the
short-range AFM spin correlations between the localized magnetic moments are
gradually increased. In order to put the present results on a more solid
ground, further investigation beyond the mean-field theory is certainly
needed.

\textbf{Acknowledgments.}

The authors would like to acknowledge the support of NSF of China.

\end{document}